\documentclass[12pt]{article}

\usepackage{amsmath}
\usepackage{amssymb}

\usepackage{graphicx}

\usepackage{cite}

\usepackage{color} 
\usepackage{units} 
\usepackage{changepage}

\usepackage[labelfont=bf,labelsep=period,justification=raggedright]{caption}

\makeatletter
\renewcommand{\@biblabel}[1]{\quad#1.}
\makeatother

\date{}

\title{To P or not to P: on the evidential nature of P-values and their place in scientific inference}

\begin{document}
\maketitle

\begin{flushleft}
\textbf{Michael J. Lew}, Department of Pharmacology and Therapeutics, University of Melbourne, Parkville, Victoria, Australia

E-mail: michaell@unimelb.edu.au
\end{flushleft}

\noindent \today

\section*{Abstract}

The customary use of P-values in scientific research has been attacked as being ill-conceived, and the utility of P-values has been derided. This paper reviews common misconceptions about P-values and their alleged deficits as indices of experimental evidence and, using an empirical exploration of the properties of P-values, documents the intimate relationship between P-values and likelihood functions. It is shown that P-values quantify experimental evidence not by their numerical value, but through the likelihood functions that they index.  Many arguments against the utility of P-values are refuted and the conclusion is drawn that P-values are useful indices of experimental evidence. The widespread use of P-values in scientific research is well justified by the actual properties of P-values, but those properties need to be more widely understood.
\\

\noindent Key words: P-value; significance test; likelihood; likelihood function; evidence; inductive inference; statistics reform.

\section{Introduction}
It is uncontroversial to say that P-values are very widely used in scientific research. For example, six of the twelve research articles and reports in the December 14 2012 issue of \textit{Science} and 20 out of 22 in the December 2012 issue of \textit{Journal of Pharmacology and Experimental Therapeutics} use P-values when describing their experimental results, specifying them either exactly or as being less than various thresholds. On the basis of such ubiquity it might be assumed that P-values are useful for scientific inference and that practicing scientists need little explanation of them. However, whether that is the case, or even \textit{should be} the case is controversial.

Significance tests and the P-values that they yield have been under attack both statisticians and non-statisticians since they first became widely used. Papers critical of them are so myriad that even a simple listing might be as long as this whole paper. Conveniently, many alleged deficiencies can be gleaned without reading beyond some titles: P-values are ``not a useful measure of evidence'' [1] and may be completely irreconcilable with evidence [2]. They ``predict the future only vaguely'' [3] and are ``impossible'' [4]. They are often confused with error rates [5], ``what they are'' is logically flawed and ``what they are not'' is coherent [6]. Significance tests are ``insignificant'' [7], non-empirical products of sorcery [8] that have been regularly ``abused and misused'' [9]. There is at least a ``dirty dozen'' of ways that P-values are regularly misinterpreted and, as  ``Even statisticians are not immune'' to those misinterpretations [10], ``you probably don't know P'' [11]. The continued use of significance tests is a ``pervasive problem'' [12] because they answer a question that no-one means to ask [13].

The previous paragraph list an apparently damning set of shortcomings that, if true and relevant, would mean that continued use of P-values for statistical support of scientific inference should not be allowed. From a practical point of view, therefore, a very important question is whether scientists choose to use significance tests and P-values in making inference because of a mistaken assumption that they have useful properties, or because they do actually have useful properties. To decide that question it is necessary to characterize those properties.

\subsection{What is a significance test and what is a P-value?}

A significance test is not a hypothesis test [11]. That will will be self-evident to many readers, but not all. Consider the likely responses by non-statistically sophisticated users of statistics to this question: which of those two types of procedure is referred to by the common phrase `null hypothesis significance test'?

A significance test yields a P-value whereas a hypothesis test yields a decision about acceptance of the null hypothesis or an alternative hypothesis. Frameworks exist that attempt to amalgamate significance and hypothesis tests [e.g. 14] or to append desirable inferential aspects of significance testing onto hypothesis testing [e.g. 15, 16] but those frameworks are controversial and have no been widely adopted. Nonetheless there is mixed approach in very widespread use. Unfortunately it is not an intentional mixture but an accidental hybrid that has been called a mishmash [17, 18], and it is a dysfunctional mishmash because the two approaches are incompatible [19, 11, 20]. The phrase `null hypothesis significance test' should be avoided because it is confusing and, arguably, is itself a product of confusion.

An essential role for P-values is a core difference between significance tests and hypothesis tests. P-values are conventionally defined with reference to the null hypothesis. For example, the author recently defined it in this way:

\begin{quote}
To be specific, a P-value obtained from an experiment represents the long-run frequency of obtaining data as extreme as the observed data, or more extreme, given that the null hypothesis is true.	[11 p. 1560]
\end{quote}

\noindent The other common style of definition specifies tail areas under sampling distributions, which amounts to the same thing. However, judging from the obvious confusion in many publications regarding the properties of P-values, neither style of definition serves well as an explanation. The introductory listing of alleged shortcomings of P-values may give the impression that confusion about P-values takes many forms but, while that may be true to a degree, one form of confusion leads more or less directly to the others. That primary confusion is that P-values measure error rates. 

The idea that P-values measure type I error rates is as pervasive as it is erroneous, and it comes hand in hand with the significance test-hypothesis test hybrid. It might be seen as a natural extension or corollary of the P-value definition quoted above and, given that many introductory level textbooks actually introduce P-values within the hybrid framework, such a misunderstanding is itself understandable. However, even though deficiencies of textbooks in that regard have been noted many times [e.g. 18, 21, 5, 11] and sometimes analyzed in depth [19, 22, 23], textbooks are not entirely to blame. It can reasonably be said that R.A. Fisher himself was a contributor to the adoption of the hybrid approach. His writings are often difficult to fathom, his approach to argument was often to `play the man rather than the ball', and even while promoting P-values as indices of evidence against the null hypotheses he advised: 

\begin{quote}
It is usual and convenient for experimenters to take 5 per cent as a standard level of significance, in the sense they are prepared to ignore all results which fail to reach this standard \hfill [24 p. 13]
\end{quote}

\noindent The dichotomization implied by that statement gives the impression that P-values fit into the error-decision framework of Neyman and Pearson. In same vein, Neyman and Pearson may also have contributed to the hybridization. They wrote:

\begin{quote}
We may accept or reject a hypothesis with varying degrees of confidence; or we may decide to remain in doubt.	\hfill	[25 pp. 295-296]
\end{quote}

\noindent That statement appears to make space for experimental conclusions other than the all-or-none decisions usually associated with their approach, but it is only an informal space: the mathematical aspects of their work leave no room for a decision to `remain in doubt'.

Real problems arise with the hybridized approach because P-values and the error rates of the Neyman--Pearsonian error decision framework are quite different. The error rates come not from the statistics \textit{per se}, but from the behavior of the experimenter upon seeing the statistics---what Neyman eventually called inductive behaviour [26]. However, the rarity of specified alternative hypotheses and sample size calculations in scientific research publications [27] and the multiple levels of $\alpha$ in statements of P $< \alpha$ without specific justification in terms of power and error tolerance make it clear that few scientists actually practice inductive behaviour.  A likely reason for the non-adoption of inductive behaviour is that it is incompatible with many scientific activities, as can be gleaned from this oft-quoted passage from Neyman and Pearson's original publication of their framework:

\begin{quote}
We are inclined to think that as far as a particular hypothesis is concerned, no test based upon the theory of probability can by itself provide any valuable evidence of the truth or falsehood of that hypothesis.

But we may look at the purpose of tests from another view-point. Without hoping to know whether each separate hypothesis is true or false, we may search for rules to govern our behaviour with regard to them, in following which we insure that, in the long run of experience, we shall not be too often wrong.	\hfill [25  pp. 290-291]
\end{quote}

In the first sentence Neyman and Pearson opine openly and explicitly that the results of a particular experiment cannot be used to discern the truth of the `particular hypothesis' of that experiment. That means that, within that framework, experimental results cannot be used as evidence for or against statements regarding the state of the world within that experiment. The quoted passage has been widely reproduced, but its implication for scientific evidence seems to be rarely enunciated. Perhaps its discordance with real scientific inference is so extreme that few who read that passage can believe that they have grasped its true meaning. Certainly it is difficult to accept the consequences of the passage, for how could the result of an experiment fail to tell the experimenter about the local state of the world? The answer is that it can do so when the experimenter is required to ignore the evidence in the results and to focus instead on the long-term error rates that would attend various behaviours.

The long-run error rates associated with an experiment are a property of the experimental design and the behaviour of the experimenter rather than of the data. The `size' of the experiment, $\alpha$, is properly set before the data is available, so it cannot be data-dependent. In contrast, the P-value from a significance test is determined by the data rather than the arbitrary setting of a threshold. It cannot logically be an error rate because it doesn't force a decision in the way that inductive behaviour does, and if a decision is made to discard the null hypothesis when a small (presumably) P-value is observed, the decision is made on the basis of the smallness of the P-value in conjunction with whatever information that the experimenter considers relevant. Thus the rate of erroneous inferences is a function of not only the P-value but the quality and availability of additional information and, sometimes, the intuition of the experimenter. P-values are not error rates, whether `observed', `obtained' or `implied'.

So, what exactly is a P-value and how should it be interpreted? Fisher regarded it as an indicator of the strength of evidence against a null hypothesis. The link between P-values and evidence is strong, but it is somewhat indirect and, as previously noted, that link is sometimes disputed or disparaged. Fisher justified the use of P-values for inductive inference by noting that a small observed P-value indicates that either an unusual event has occurred or the null hypothesis is false. That is sometimes called Fisher's disjunction and it implies that an experiment casts doubt on the null hypothesis in some sort of proportion to the smallness of the observed P-value. Obviously, for practical application, it is desirable to be able to specify the relationship between P-values and evidence more completely than a vague phrase like `in some sort of proportion' and the next section of this paper explores that relationship empirically. It is hoped that full documentation of those properties will not only help to reduce the misapprehension that P-values are error rates, but also encourage a more thoughtful approach to the evaluation of experimental results.

\section{The distribution of P-values  and likelihood}

 I take as a starting point the fact that P-values are data-dependent random variables [31] and the view that likelihood functions encapsulate the evidential aspects of data [28-30]. It is not intended to discard the conventional definitions of P-values, but to gain insights into P-value-ness---perhaps different from those afforded by those definitions---through empirical exploration. The working definition of a P-value in this section is `that value returned by the R function t.test'. Student's \textit{t}-test for independent samples was chosen as the exemplar significance test for its ubiquity, and while it is possible that some of the properties documented will be specific to that test, most will certainly be general.
 
Simple Monte Carlo simulations allow the exploration of P-values. Figure 1 shows the two-tailed P-values resulting from one million Student's \textit{t}-tests for independent groups of $n = 10$ (i.e. 18 degrees of freedom) with the true difference between the groups being a uniform random deviate in the range $-4$ to $4$ times the population standard deviation. The resulting cloud of P-values has mirror-symmetry around effect size of zero, with the P-values increasingly clustered towards the x-axis as the absolute effect size increases. (It is notable, and perhaps useful for pedagogic purposes, that there is no sudden change in the distribution of P-values at any level and so there is not a `natural' place to set a threshold for dichotomizing the P-values into significant and not significant.) 

That cloud of P-values is informative only in a qualitative fashion, but the distribution of P-values can easily be quantified using the probability density functions obtained as vertical density sections through the cloud (Figure 2). Probability density functions of P-values are rarely seen, with the consequence that many users of statistical tests are unfamiliar with the patterns of P-values that can be expected in various experimental circumstances [3, 32]. Those patterns show that where the null hypothesis is false, a small P-value is more often observed than a large P-value and anything that would increase the power of an experiment (e.g. larger effect size or sample size) leads to the P-value probability density function being increasingly piled up at the left hand end of the graph.

\begin{figure}[!ht]
\begin{center}
\includegraphics[width=0.5\linewidth]{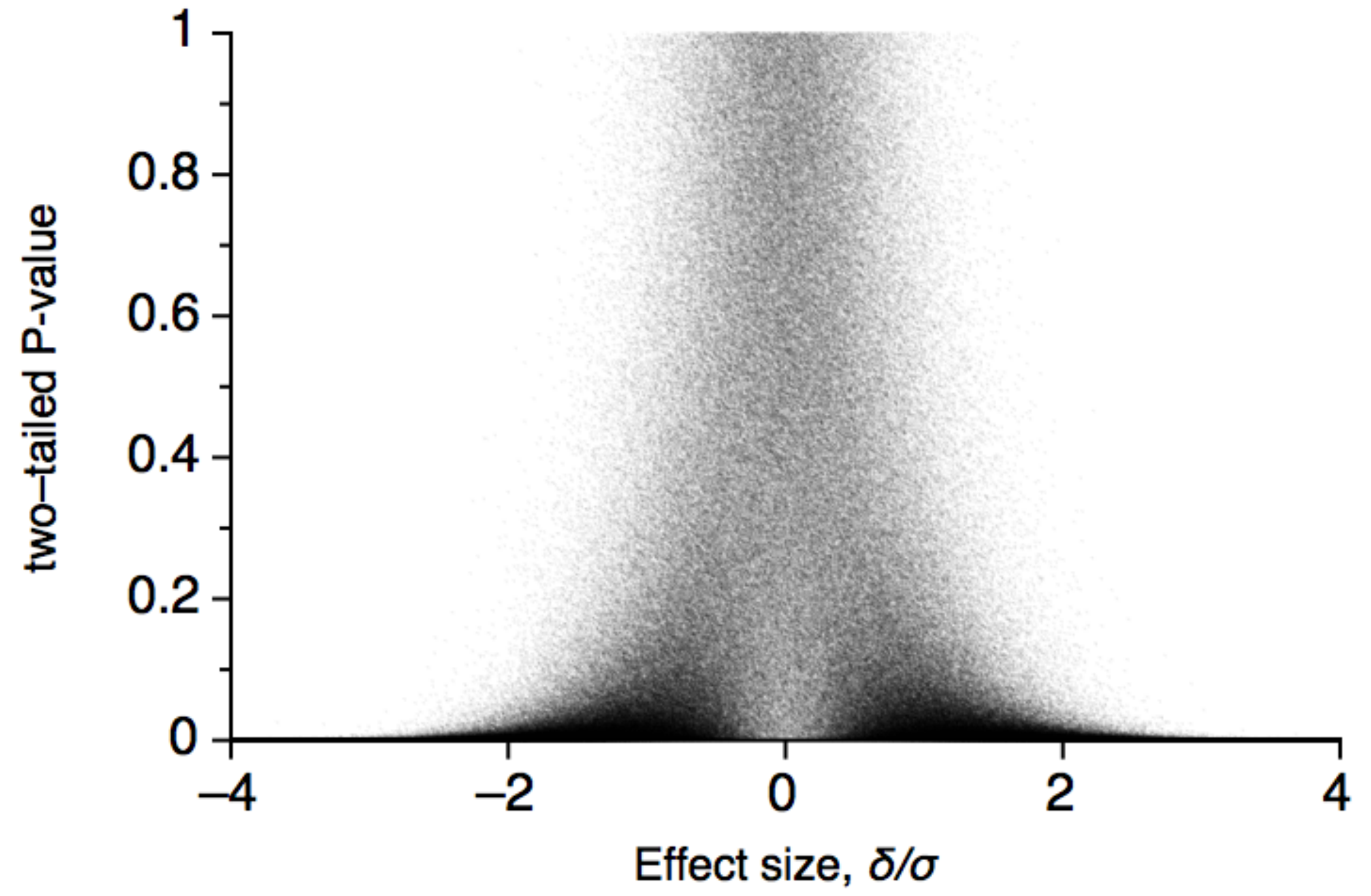}
\end{center}
\caption{
Distribution of two-tailed P-values from Student's \textit{t}-test for independent samples ($n=10$ per group). Data are the results from $10^6$ Monte Carlo simulations where the true effect size was a uniform random variate between -4 and 4 times the population standard deviation.}
\label{Fig_1}
\end{figure}

\begin{figure}[!ht]
\begin{center}
\includegraphics[width=0.5\linewidth]{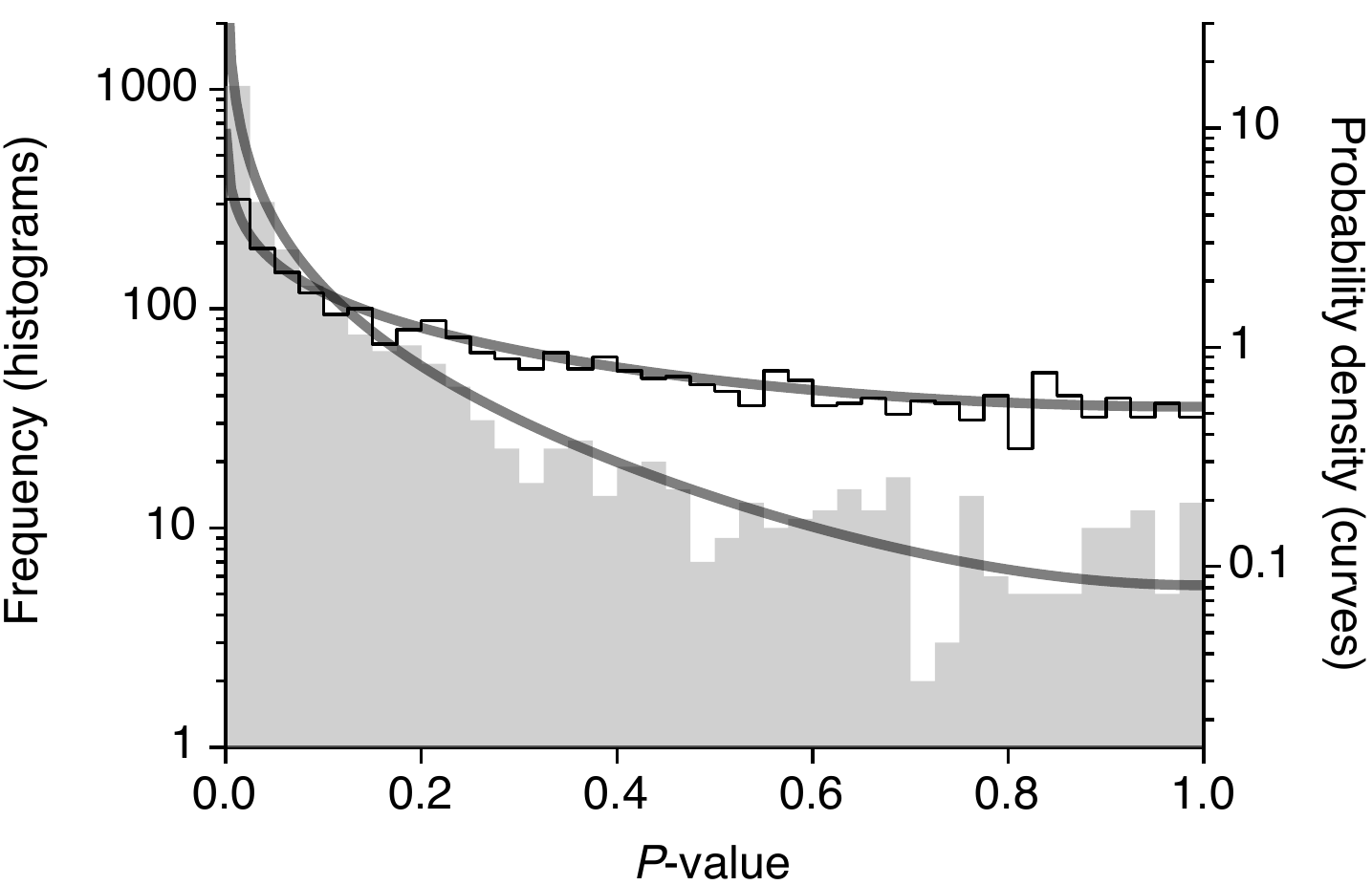}
\end{center}
\caption{
	Probability density functions for two-tailed P-values from Student's \textit{t}-test for independent samples (curves) and frequency histograms of the P-values in the simulations from effect sizes, \nicefrac{$\delta$}{$\sigma$}, in the range of 0.495--0.505 (lines) and 0.995--1.005 (shaded). The histograms are effectively vertical sections of the blackness in Figure 1, and the density functions were obtained using the R code given in the appendix.}
\label{Fig_2}
\end{figure}


P-value probability density functions, like those in Figure 2, have utility in illustrating the properties of P-values and, while they will be unsurprising to many, they do have some pedagogical utility particularly when combined with the median P-values to provide an alternative to the conventional power calculations that are part of the Neyman--Pearson framework [31, 33]. However, they are not relevant to interpretation of an observed P-value that is already in hand because each distribution is specific to an effect size, and it is usually the size of the effect, or its existence, that is the subject of the experiment. Thus P-value probability density functions, like power calculations [34], can be useful in the planning stage but have at most tangential relevance to interpretation of the actual results of an experiment. A different picture emerges if, instead of vertical sections through the P-value cloud of Figure 1, horizontal sections are taken at the level of the observed P-value. The horizontal sections tell us how likely any effect size is to yield the observed P-value, information that is directly useful for interpreting experimental results. The horizontal sections are likelihood functions.

\subsection{Likelihood functions from P-values}

Mathematical likelihood is a relatively rare thing in statistics in that it means almost exactly what a non-statistician would expect. The simplicity of likelihood can be seen in Figure 1: for any observed P-value, the real effect size is more likely to have corresponded to the darker region of the graph than the lighter regions. However,  despite its apparent simplicity, likelihood is not prominently featured in most introductory statistics textbooks, and it is often completely absent. It is not well enough known that I can proceed without some definition and explanation.

The first thing to note is that likelihoods relate to hypothesised parameter values rather than the observations so it is correct to speak of the likelihood of the effect sizes in Figure 1 rather than the likelihood of the observed P-values. The likelihood of a particular hypothesised value of the effect size, say $\nicefrac{\delta}{\sigma} = 2.4$, is proportional to the probability of the observation under the assumption that that true effect size is $2.4$. If we use $\theta$ to stand for the set of all possible values of $\nicefrac{\delta}{\sigma}$ and say that the observation is a P-value of 0.01, then the likelihood function of $\theta$ is

$$ L(\theta) \propto Pr(\text{P}=0.01 \,|\,\theta). $$

\noindent The reason that likelihood can be defined only up to a proportionality constant is that the probability of an observation is affected by the precision of the observation. For example, the probability of observing a P-value that rounds to 0.01 will necessarily be higher than the probability of observing P=0.010, or P=0.0100 and so on. In some circumstances the existence of an unknown proportionality constant limits the utility of likelihood for comparisons between disparate systems which would usually have different constants, but in the vast majority of cases it is only necessary to compare likelihoods within a single likelihood function and so there is a single shared constant which can be cancelled out in a ratio of likelihoods.

Likelihoods can be used is as measures of `support' provided by the observed data for hypothesised parameter values. It probably seems natural to most readers that a small observed P-value would support a hypothesis of effect size greater than zero more strongly than it would support the hypothesis of effect size equals zero---it is the ratio of the likelihoods that provides a scaling of the relative support. Thus the likelihood function shows that hypothesized effect sizes corresponding to the darker regions are better supported by the observed P-value than those corresponding to the lighter regions. As Royall puts it, ``The law of likelihood asserts that the hypothesis that is better supported is the one that did a better job of predicting what happened'' [35]. The law of likelihood says that an observation $ x $ gives support for a hypothesis that predicts $ x $ with probability $p_1$ over a rival hypothesis that predicts $ x $ with probability $p_2$ to the degree $\nicefrac{p_1}{p_2}$. To connect that to Figure 1, consider that we have a continuum of hypotheses, each proposing a different effect size. The likelihood that any particular effect size is equal to the true effect size is directly related to the blackness of the horizontal slice of Figure 1 corresponding to the observed P-value, and the relative support for any two hypotheses is proportional to the ratio of their likelihoods.

Calculation of the relevant likelihood functions can be achieved using the non-central Student's \textit{t} distribution, but in the context of an exploration of P-values it is more instructive to calculate them via power curves as functions of effect size. Let $\theta$ be the set of values that can be taken by the effect size, $\nicefrac{\delta}{\sigma}$, then for a particular sample size we can define the power curve as a function of effect size: 

\begin{equation}
\label{powerFn1}
f(\theta) = Pr(\mathrm{P} \leq \alpha \, |\, \theta), \quad \theta \neq 0.
\end{equation}

\noindent The specification that $\theta \neq 0$ is only necessary because power is undefined where the null hypothesis is true and so, as we are not concerned with error rates, it can be omitted. After rewriting the power curve as an integral our route to the likelihood function becomes clear.

\begin{equation}
\label{powerFn2}
f(\theta) =\int_{x=0}^{x=\alpha}\! Pr(\mathrm{P}=x\,|\,\theta) \,dx.
\end{equation}

\noindent The likelihood curve is also a function of $\theta$:

\begin{equation}
\label{LikeFn1}
L(\theta) \propto Pr(\mathrm{P} = x \, |\, \theta).
\end{equation}

\noindent That function is clearly the first derivative of that power curve in equation \ref{powerFn2} with respect to $x$, which is itself on the scale of P:

\begin{equation}
\label{LikeFn2}
L(\theta) \propto \frac{df}{dx}(x).
\end{equation}

\noindent That likelihood function can conveniently be calculated using the power.t.test() function of R using the code in the appendix.

\subsubsection{Likelihood functions from two-tailed P-values}
Figure 3 shows how that blackness varies with the effect size at the level of P=0.01, along with the corresponding likelihood function. Interpretation of the likelihood function in Figure 3 is quite straightforward: the higher the likelihood at a given effect size, the higher the support given by the observed P-value for that effect size. Thus the likelihood function shows the support of any value of \nicefrac{$\delta$}{$\sigma$} relative to any other. For example, an observed P-value of 0.01 supports the effect size of $\nicefrac{\delta}{\sigma} = 1.5$ about 17 times more strongly than it supports $\nicefrac{\delta}{\sigma} = 0$. The support quantified by the function relates not only to the null hypothesis, so it simultaneously shows that the observed P-value of 0.01 supports an effect size of $\nicefrac{\delta}{\sigma} = 2$ about six times more than it supports \nicefrac{$\delta$}{$\sigma$} = 0, but nearly three times less than it supports $\nicefrac{\delta}{\sigma} =  1.5$.

\begin{figure}[!ht]
\begin{center}
\includegraphics[width=0.5\linewidth]{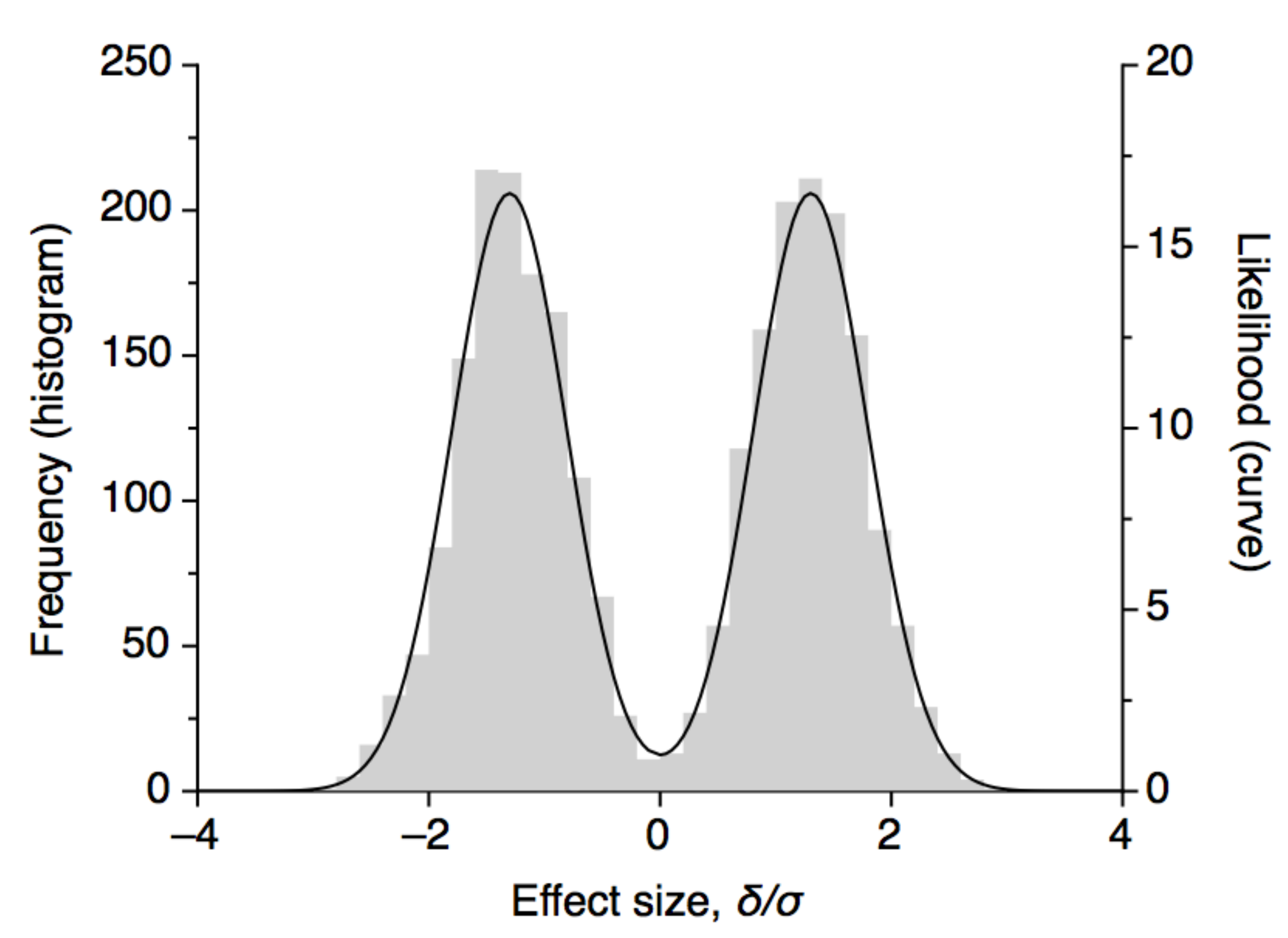}
\end{center}
\caption{
	Likelihood function for a two-tailed P-value of 0.01 (curve) and the frequency of occurrence of P-values in the range of 0.0095--0.0105 at all effect sizes in the simulations (histogram). The histogram is effectively a horizontal section of the blackness of Figure 1, and the likelihood function was obtained using the R code given in the appendix.}
\label{Fig_3}
\end{figure}

Those interpretations of the likelihood functions are both convenient and easy, but there appears to be a large fly in this ointment: the likelihood function in Figure 3 suggests that negative and positive values of $\nicefrac{\delta}{\sigma}$ are supported equally well, so if the only thing known about an experimental result was the P-value then likelihood functions like that in Figure 3 would yield ambiguous results. In a real experiment the two-tailed P-value is always accompanied by knowledge of the direction of the observed effect, so no investigator is going to credit positive and negative effects equally, but that problem does indicate that the likelihood function from two-tailed P-values is not ideally suited to quantification of experimental evidence. We do not have to give up on likelihood functions, though, because the problem lies in the use of effect-direction-agnostic two-tailed P-values. If we utilize one-tailed P-values we end up with an unambiguous likelihood function.

\subsubsection{Likelihood functions from one-tailed P-values}

One-tailed P-values from a Student's \textit{t}-test are not, as sometimes asserted, half of the equivalent two-tailed P-value---if that was the case then a one-tailed P-value could never be larger than 0.5. Instead, they depend on the sign of the test statistic. When \textit{t} is positive then the one-tailed P-values are, indeed, half of the two-tailed, but when \textit{t} is negative the one-tailed P-value is 1 minus half of the two-tailed P-value. Figure 4 shows the distribution of one-tailed P-values.

  \begin{figure}[!ht]
\begin{center}
\includegraphics[width=0.75\linewidth]{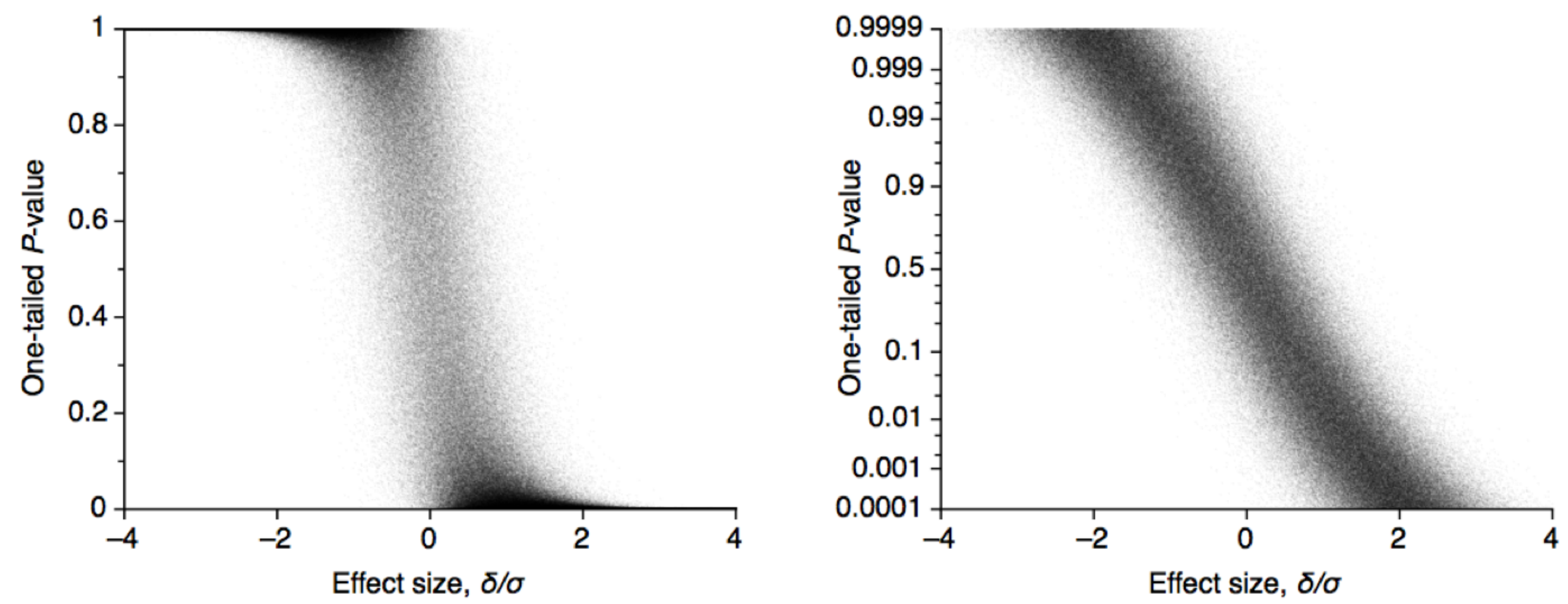}
\end{center}
\caption{
	Distribution of one-tailed P-values from Student's \textit{t}-test for independent samples ($n=10$ per group). Data are the results from $10^6$ Monte Carlo simulations where the true effect size was a uniform random variate between -4 and 4 times the population standard deviation. In the panel on the right a non-linear vertical axis is used for improved clarity.}
\label{Fig_4}
\end{figure}

As before, the probability density functions for one-tailed P-values and likelihood functions can be visualized as vertical and horizontal sections through the cloud of P-values in Figure 4. The P-value probability density functions are heavy at the left or right hand end, depending on the sign of the observed effect (Figure 5), and the likelihood functions are unimodal (Figure 6).

If it is intended to use a P-value as the sole index of the evidence provided by the observed data, then the fact that two-tailed P-values imply bimodal likelihood functions is a serious problem---an index of evidence that points in both directions simultaneously is deficient in specificity. Thus, a conclusion that follows from these investigations is that one-tailed P-values are better evidential indices than are two-tailed P-values, and one-tailed P-values should be used whenever an experimenter intends to use them as an indices of evidence. One-tailed P-values are controversial, so arguments for and against their use will be discussed before returning to the general issue of P-values and experimental evidence.

 \begin{figure}[!ht]
\begin{center}
\includegraphics[width=0.5\linewidth]{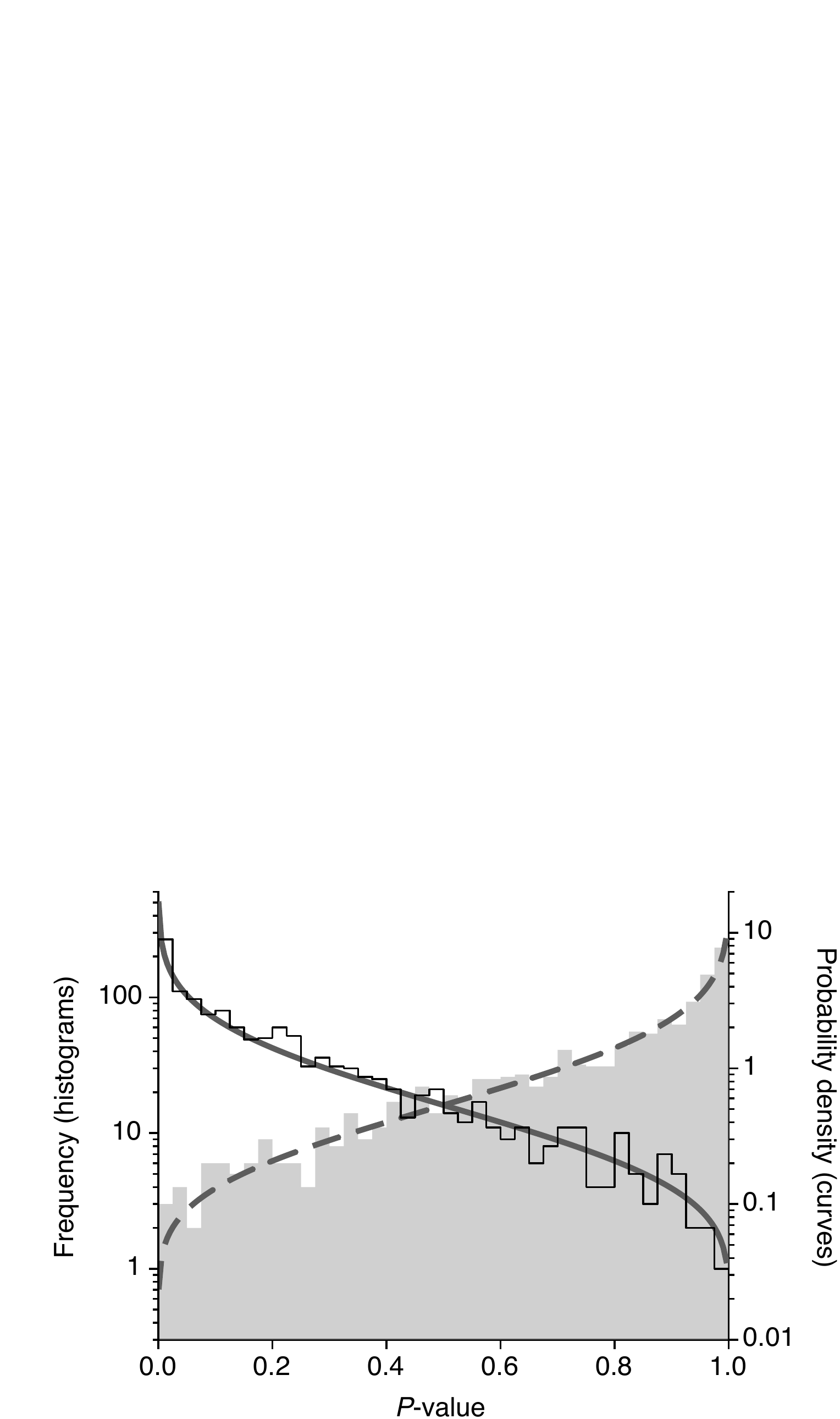}
\end{center}
\caption{
	Probability density functions for one-tailed P-values from Student's \textit{t}-test for independent samples (curves) and frequency histograms of the P-values in the simulations from effect size in the range of -0.495 to -0.505 (shaded) and 0.495 to 0.505 (lines). The histograms are effectively vertical sections of the blackness in Figure 4, and the density functions were obtained using the R code given in the appendix.}
\label{Fig_5}
\end{figure}

 \begin{figure}[!ht]
\begin{center}
\includegraphics[width=0.5\linewidth]{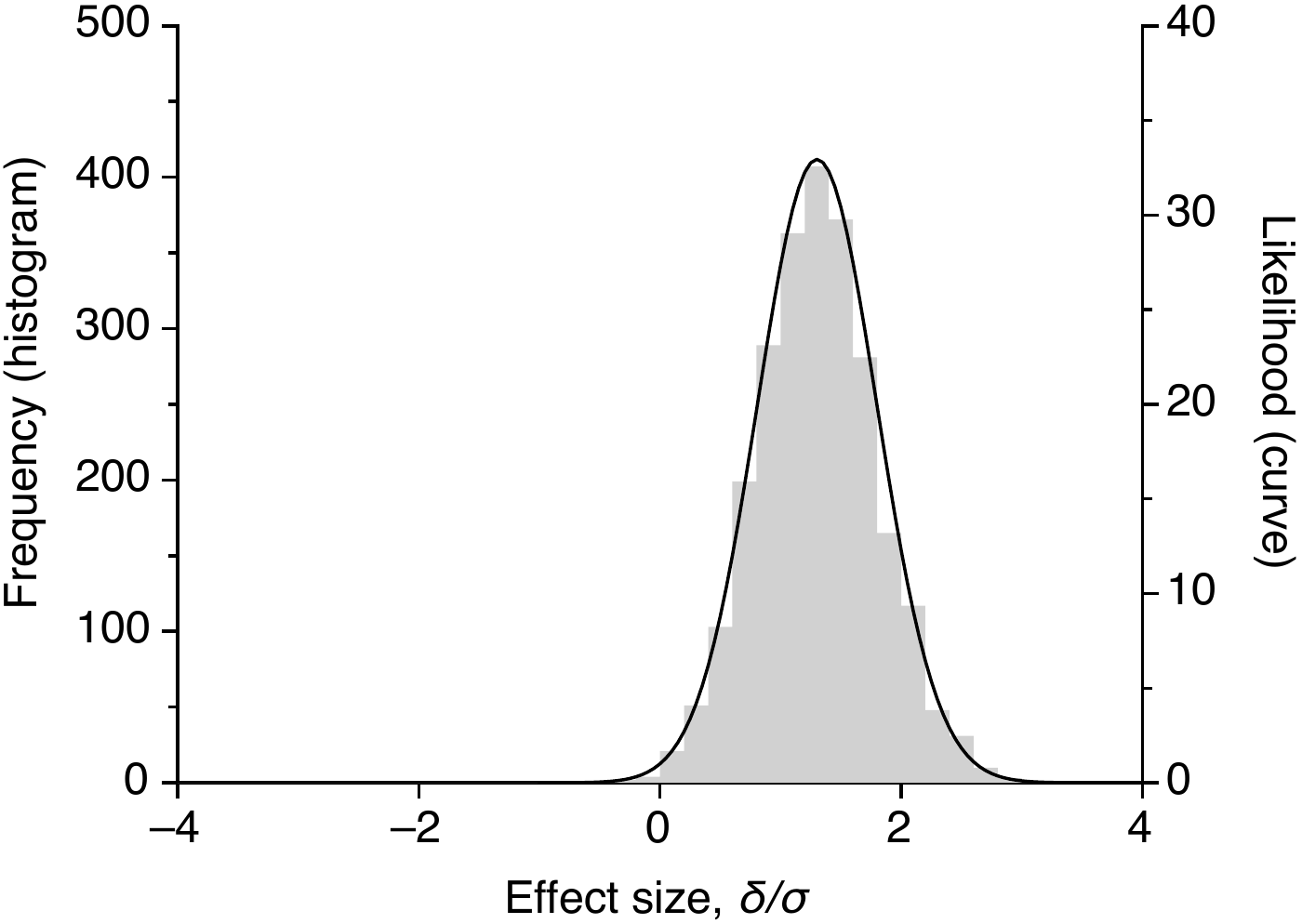}
\end{center}
\caption{
Likelihood function for a one-tailed P-value of 0.005 (curve) and the frequency of occurrence of P-values in the range from 0.004975 to 0.00525 at all effect sizes in the simulations (histogram). The histogram is effectively a horizontal section of the blackness of Figure 4, and the likelihood function was obtained using the R code given in the appendix.}
\label{Fig_6}
\end{figure}

\section{Use one-tailed P-values (or two-tailed)}

While statistical textbooks are not always helpful regarding the choice between one- and two-tailed testing [36], there are many commentary and review papers extolling the virtues of one-tailed P-values [e.g. 37, 38, 39]. However, that is far from a consensus position because an even larger set of papers argue that one-tailed tests should \textit{almost never} be used [e.g. 36, 40, 41]. The tone of the discussion can be seen in a short note by the famous psychologist H.J. Eysenck that appeared to conclude a long-running debate about the merits of using one-tailed P-values in psychological research:

\begin{quote}
A statement of one-tailed probability is not a statement of fact, but of opinion, and should not be offered instead of, but only in addition to, the factual two-tailed probability.	\hfill[42 p.270]

\end{quote}

\noindent However, users of one-tailed P-values can take support from no less an authority than `Student' (W.S. Gossett) who used one-tailed P-value to conclude that \textit{l}-hyoscyamine bromide was a soporific in the original description of his eponymous \textit{t}-test [43].

It is notable that virtually all papers discussing the appropriateness of one-tailed P-values do so implicitly or explicitly from within the Neyman--Pearson paradigm. For example, Freedman is explicit:

\begin{quote}
This article is concerned only with the hypothesis-testing paradigm; p-values are used in this article only as an operational means for deciding whether or not to reject the null hypothesis and not as measures of evidence against the null hypothesis.		\hfill	 [44]
\end{quote}

\noindent However, many other papers expounding about the number of tails that should adorn a P-value are silent about whether they are considering significance tests and hypothesis tests. That is not an idle observation, because conclusions about the suitability of one- and two-tailed P-values for statistical inference will necessarily be influenced by the framework within which the issues are considered. In this section it will be shown that arguments about the applicability of one-tailed P-values to inductive behaviour within an error-decision framework can be irrelevant to the utility of those P-values as evidence.

\subsection{Effects in an unexpected direction}

An important argument against the use of one-tailed tests is that they force the experimenter who observes an effect in the unexpected direction to choose between ignoring the effect and `cheating' by pretending that the direction was expected all along. The following is on one of the slides that the author has used in teaching the basics of statistical reasoning and testing to young biomedical researchers:

\begin{quote}
One-sided tests are always more powerful than two-sided, but are you prepared to ignore a result where your drug makes the responses a lot smaller? Tails: just use two!
\end{quote}

\noindent (That slide has now been ammended!) 

If one-tailed tests imply that one has to ignore effects in an unexpected direction then their use would require a ``lofty indifference to experimental surprises'' [45]. That sounds like a compelling reason to avoid one-sided tests, and it is widely accepted as such. However, it is also unrealistic. Lombardi \& Hurlbert suggest that in practice few, if any, clear effects in an unexpected direction will be equated with no effect:

\begin{quote}
In every \textit{d} [effect size] associated with a low P value, regardless of sign, there is a good story. And we have never known a colleague who shirked at its telling. Such behaviour is predictable.\hfill	[36 p. 453]
\end{quote}

\noindent Furthermore, such behaviour is desirable where the data are viewed in light of the evidence that they contain. A large effect in the unexpected direction yields a one-tailed P-value that is not large and therefore to be ignored, but extreme and thus noteworthy. A one-tailed P-value of 0.995 is just as extreme as a one-tailed P-value of 0.005 and their likelihood functions are located symmetrically on either side of zero (or whatever value is chosen as the null hypothesis) and thus they offer equally strong evidence against the null hypothesis.  From the standpoint of evidence, the issue of ignoring unexpected effects is entirely moot.

An experimental result in the unexpected direction might provide unexpected evidence, but it nonetheless provides evidence, and while the predicted-ness of an outcome should properly affect how an experimenter thinks about the result, but it does not affect the evidence. Thus the ``lofty indifference'' argument against one-tailed P-values disappears as soon as one discards the notion that P-values are something to do with error rates.

\subsection{Maintaining `standards'}

Other reasons for not preferring one-tailed P-values include the notion that adoption of one-tailed P-values entails a lowering of the standards for publication of results, with an attendant risk of more papers containing unreproducible results and erroneous conclusions. It is true that one-tailed P-value of 0.05 requires less extreme results than a two-tailed P-value of 0.05---the data that gave that one-tailed result would give a one-tailed P-value of 0.1---but the simple interconvertability between one- and two-tailed P-values leads others to argue that the choice of tails is unimportant, as long as their number is clearly specified and the exact P-value given. Some who might forbid the use of one-tailed P-values may not be mollified by that argument because their concerns often relate not to the P-values \textit{per se}, but to the assumed tendency for attainment of `statistical significance' to be taken as sufficient reason to `believe' a result (or, at least, to publish it). In light of that assumption, to license the use of one-tailed P-value would be tantamount to a halving of the protection against false positive claims. However it is the tendency to uncritically believe that a `significant' result is `real' (or publishable) that is at the heart of this problem. The credibility gained from the words `statistically significant' is at fault rather than the nature of one-tailed P-values. If we wish to reduce publication of unreliable inferences based on weak evidence then we should assess the evidence rather than relying on mindless protection by a rickety hurdle of error rate-related `significance' [11, 46].

\subsection{An equivocal recommendation}

So, how many tails do I recommend scientists use? The short answer is one, but the longer answer will make it clear that the number of tails doesn't much matter. All arguments for two tails that are only relevant to the Neyman--Pearsonian error-decision framework can safely be ignored when working within an evidential framework, but, at the moment, statistically non-sophisticated readers will not understand that a one-tailed P = 0.99 offers as much evidence for a non-zero effect as P = 0.01, and so there is a substantial risk of misinterpretation associated with one-tailed P-values for unexpected observations. Conveniently, that issue can be circumvented by using two-tailed P-values whenever it is convenient or desirable, as long as the number of tails is clearly indicated. Those two-tailed values can be converted into one-tailed values of either tail by anyone who cares to do so. Further, because the relationship between a P-value and a likelihood function is one value to one continuous function, it is a relationship of specification rather than measurement---we could simply choose to let a two-tailed P-value stand as specification of the likelihood function of the relevant one-tailed P-value if that would help. That way we can have our cake and eat it too: it doesn't matter how many tails one uses as long as the effect direction and the number of tails are specified to the reader for every P-value reported.

\section{Arguments against the use of P-values as evidence}

There are good reasons to accept Fisher's notion that P-values are a summary of the evidential meaning of experimental data analyzed by a significance test. However, so many arguments have been put forward that P-values should not be interpreted in that manner that some discussion is needed. The arguments directly addressed are listed here:

\begin{itemize}
\item P-values do not offer a consistent scaling of evidence. (P-values are affected by sample size;  P-values are affected by stopping rules.)

\item 	P-values are the product of a logically flawed approach. (Fisher's disjunction is false; P-values depend on data that were not observed; the null hypothesis is often known to be false before the experiment.)

\item P-values overstate the strength of evidence. (P is smaller than another statistical variable; a large enough sample will always yield a small P-value)

\item P-values conflict with the likelihood principle.

\end{itemize}

\subsection{Inconsistent scaling of evidence}

For P-values to be used as indices or summaries of the evidential meaning of experimental data, it is desirable that they be readily interpreted. If there was a simple and consistent relationship between the numerical value of P and evidence then interpretation of P would be trivial, but the relationship is neither simple nor consistent. However, as will be shown below, the complexity comes from the nature of evidence rather than from the nature of P-values.

\subsubsection{P-values are affected by sample size}
\label{PSampleSize}

A common criticism of P-values as indices of evidence is that their evidential meaning varies with sample size [48], a criticism that boils down to the fact that it is difficult to answer questions of this form: ``Is a result of P = 0.025 from $n = 3$ the same strength of evidence against the null hypothesis as a result of P = 0.025 from $n = 100$?'' To answer that question empirically, sets of 100,000 one-sided P-values were generated from Student's \textit{t}-test for independent samples by Monte Carlo simulation, as before, with sample sizes of $n = 3, 5, 10$ and 100. The results show that as $n$ increases, the cloud of P-values becomes narrower and steeper (Figure 7) with the consequence that, for any given P-value, the likelihood function gets narrower and closer to the null hypothesis effect size (Figure 8). Thus it is true to say that the strength of evidence summarized by a P-values varies with sample size, but that does not mean that the P-value is not a useful index, or summary, of evidence---it simply means that a P-value should always be accompanied by the sample size. The P-value and sample size together correspond to a unique likelihood function, and thus act as a summary of that function and the evidence quantified by that function.

 \begin{figure}[!ht]
\begin{center}
\includegraphics[width=0.75\linewidth]{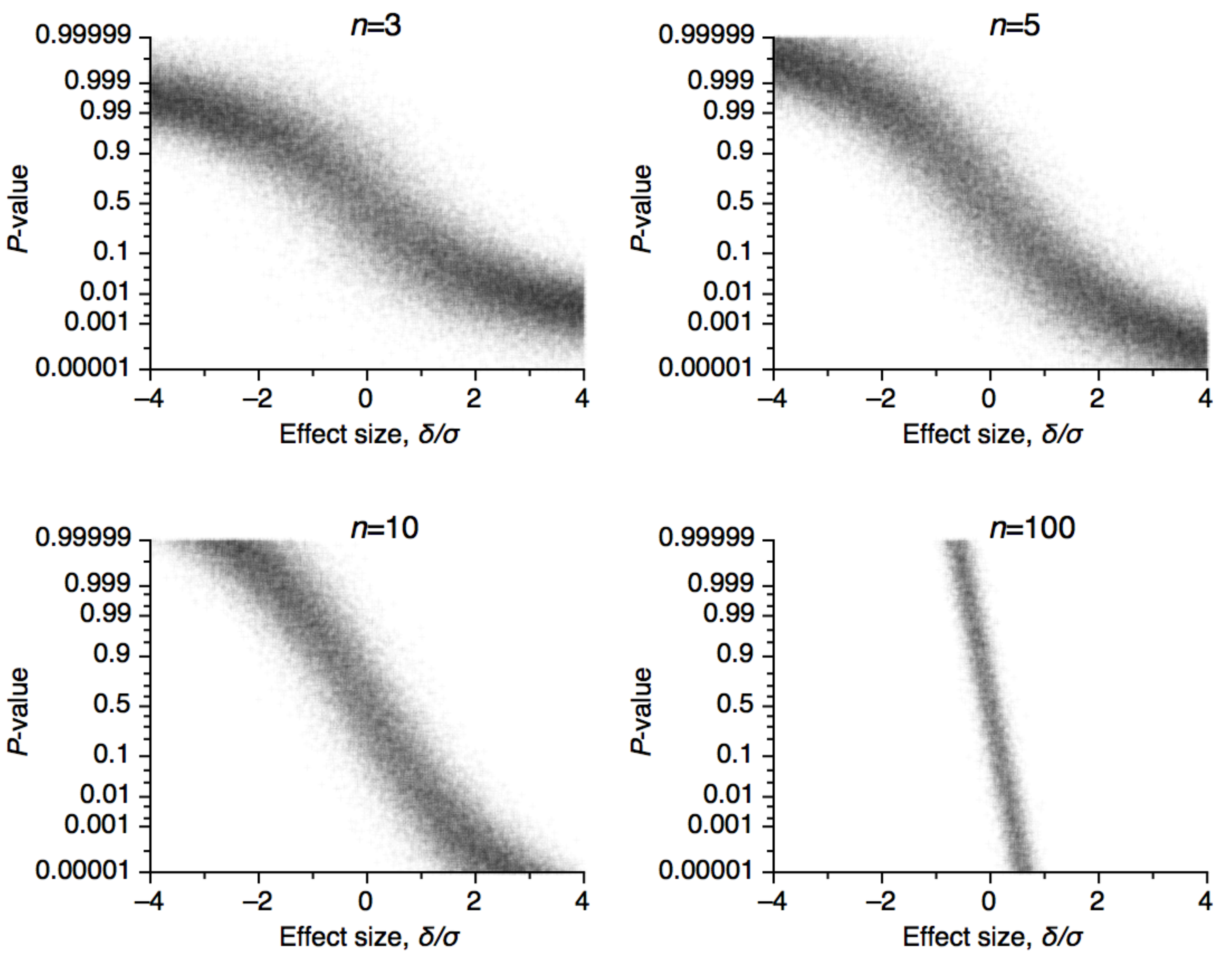}
\end{center}
\caption{
Distribution of one-tailed P-values from Student's\textit{t}-test for independent samples with various sample sizes. Data are the results from $10^4$ Monte Carlo simulations where the true effect size was a uniform random variate between -4 and 4 times the population standard deviation. A non-linear vertical axis is used for clarity.}
\label{Fig_7}
\end{figure}

\begin{figure}[!ht]
\begin{center}
\includegraphics[width=0.35\linewidth]{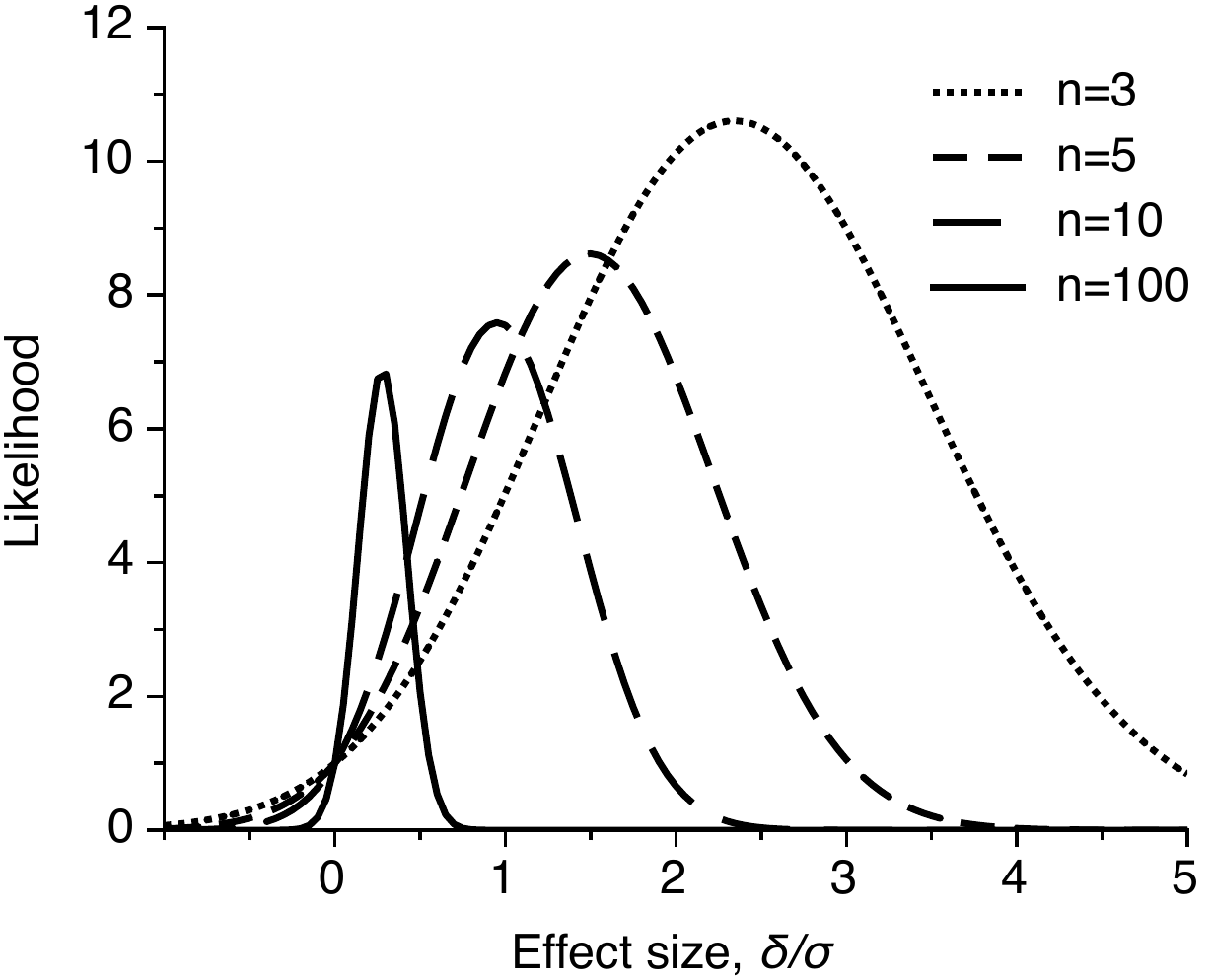}
\end{center}
\caption{
Likelihood functions for observe P-value of 0.025 with various sample sizes.}
\label{Fig_8}
\end{figure}

The difficulty here is not really in the meaning of the P-value, but in equating the evidential meaning of the same P-value obtained from different sample sizes. The question posed in the previous paragraph only asks about evidence relative to the null hypothesis, and is restricted to the dimension `strength'. While that question sounds reasonable given that the evidential nature of a P-value is usually described as being `the strength of evidence against the null hypothesis', it is actually ill-posed. Evidence encoded by a P-value and its corresponding likelihood function is not applicable only to the null hypothesis, and it is not fully specified by strength. It is better to think of it as having (at least) the two dimensions, strength and specificity, with strength relating to the height of the likelihood function and the specificity to the width and location of that function. In that way, significance testing can be thought of as being a process for estimation of the parameter on a continuous scale, rather than as a decision process for dichotomously choosing between hypotheses. 
It is reassuring to find myself in agreement with both E.T. Jaynes and R.A. Fisher. Jaynes, a leading proponent of Bayesian approaches and no friend of significance testing, said that

\begin{quote}
the distinction between significance testing and estimation is artificial and of doubtful value in statistics	\hfill	[13 p. 629]
\end{quote}

\noindent And Fisher said:

\begin{quote}
It may be added that in the theory of estimation we consider a continuum of hypotheses each eligible as null hypothesis, and it is the aggregate of frequencies calculated from each possibility in turn as true [\dots] which supply the likelihood function \hfill		[49 p. 73]
\end{quote}

\noindent Focusing on solely the strength of the evidence is insufficient because while the evidence may be \textit{against} the null hypothesis, it is also \textit{in favor} of parameter values near to the observed estimate.

The ill-posed question asked above can be re-posed in a form that is compatible with estimation and the multiple dimensions of evidence: ``What are the evidential meanings of observed P = 0.025 from samples of $n = 100$ and $n = 3$?'‘ After reference to Figure 8, that question can be usefully answered like this: the $n= 100$ result indicates that the true effect size is unlikely to be as low as zero, it is very likely to be less than 1, and can be expected to be quite close to the observed effect size; the $n= 3$ result is evidence that the true effect size is unlikely to be as low as zero but might be quite different from the observed effect size. 

The general features for interpretation of P-values that can be gleaned from those examples will not surprise anyone who has experience with P-values: smaller P-values offer stronger evidence against the null hypothesis. The meaning of a P-value is affected by the sample size, exactly as any index of evidence \textit{should} be: larger sample sizes provide higher specificity from a more reliable estimate of the true effect size.

\subsubsection{P-values are affected by stopping rules}
\label{sec:stopping_rules}

Another important criticism of P-values as indices of evidence is the notion that P-values are affected by stopping rules, and thus by the experimenter's intentions [e.g. 4]. However, that seems only to be the case when the method of calculation of the P-values is adjusted to account for the stopping rules. Such adjustments typically ensure a uniform distribution of P under the null hypothesis, a distribution that is often assumed or implied, and sometimes stated, to be required for a P-value to be valid. However, the only reason that P-values would need to be uniformly distributed under the null hypothesis is to allow them to comply with the \textit{frequentist} or \textit{repeated sampling principle}:

\begin{quote}
In repeated practical use of a statistical procedure, the long run average error should not be greater than (and ideally should equal) the long-run reported error. \hfill	[14]
\end{quote}

\noindent However P-values are not error rates. They are outside the scope of that principle and so any criticism of the utility of P-values based on non-compliance with the frequentist principle entirely moot. 

I assert that the evidential meaning of P-values are immune to the influence of stopping rules, just as likelihood functions are. That assertion is likely to elicit strong disagreement from some readers, but its truth can be demonstrated empirically by simulations with non-standard stopping rules. To that end, experiments were run with sample size $n$ = 5 and a Student's \textit{t}-test performed. In runs where the observed P-value was less than 0.05, or greater than 0.15 the P-value was accepted and the experiment stopped, but in all runs where the original P-value was between 0.05 and 0.15 an extra 5 observations was added to each group and a new P-value calculated. That two-stage stopping rule contains an interim analysis (`data peeking'), and thus has an informal sequential design. (It is worth noting, as an aside, that such features are thought to be quite common, but usually occult, in some basic biomedical research areas. `Abuses' of statistical processes are only to be expected in the circumstance where the importance of stopping rules to the resulting error rates from Neyman--Pearson hypothesis testing is rarely presented in introductory statistics textbooks, and where the description of experimental protocols in research publications have become extremely abbreviated and formulaic.)

\begin{figure}[!ht]
\begin{center}
\includegraphics[width=0.85\linewidth]{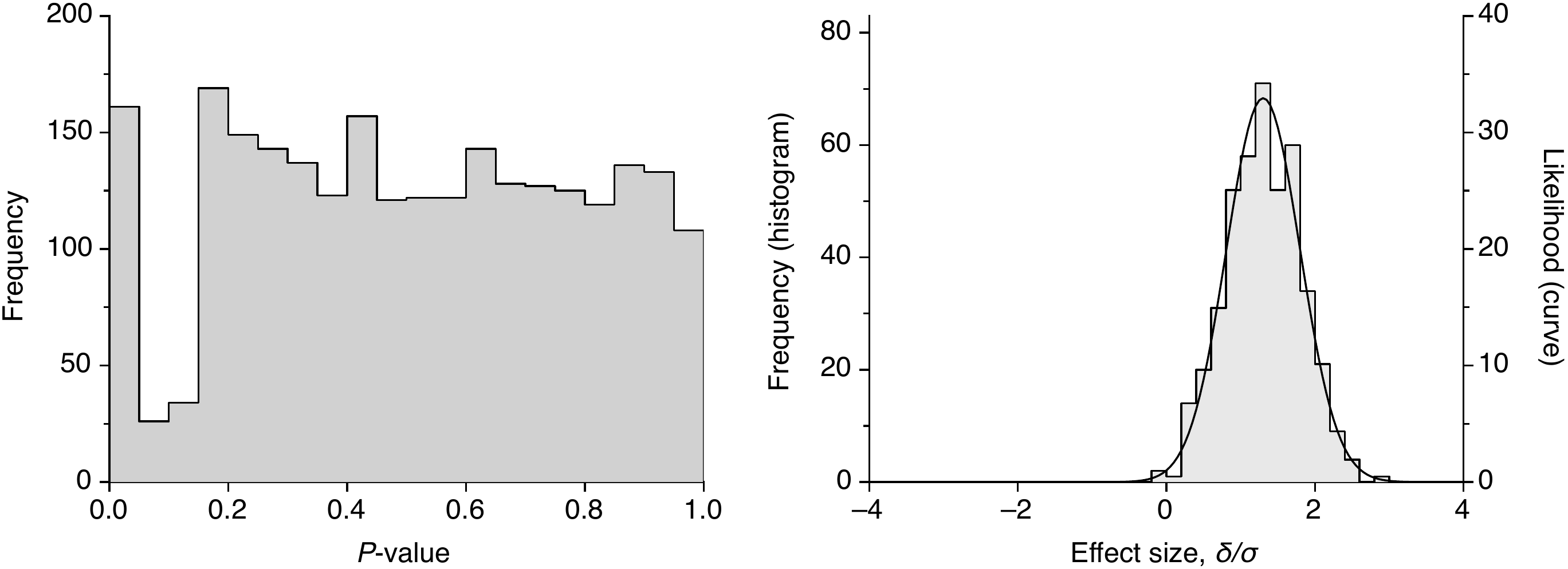}
\end{center}
\caption{
Left: Distribution of the P-values among simulations where the effect size was close to zero (-0.001 to 0.001) using the stopping rule with data peeking detailed in the text. Right: Likelihood function for a one-tailed P-value of 0.005 from $n=10$ (curve) and the frequency of occurrence of P-values in the range of 0.004975--0.00525 at all effect sizes in the simulations (histogram). (Compare the right panel with Figure 6.)}
\label{Fig_9}
\end{figure}

The results of those simulations with the null hypothesis true show a non-uniform distribution of P-values calculated without `correction' for the two-stage stopping rule, as expected (Figure 9). That non-uniformity does mean that those P-values cannot be correctly interpreted as the probability under the null hypothesis of obtaining a result at least as discrepant as that observed---the probability of obtaining a P-value less than, for example, 0.15 was substantially lower than 0.15---but that does not mean that the evidential meaning of the P-values is changed. If that were the case then we would expect to see a change in the relationship between the observed P-value and the likelihood function, which is well known to be independent of stopping rules. There is no such change, as can be seen in the second panel of Figure 9. The frequency distribution of effect sizes among the simulation runs that went on to yield P = 0.005 at $n = 10$ matches the likelihood function for P = 0.005 at $n = 10$ from a standard one-stage stopping rule. That result indicates that, despite the non-uniform distribution of P-values under the null hypothesis, the two-stage stopping rule has not affected the distribution of the P-values among the experiments that went to the second stage. (It is not necessary to demonstrate that the two-stage protocol fails to affect the runs that terminated at the first stage, because the mere possibility of extra observations that didn't actually eventuate cannot influence their P-values.) Thus the relationship between P-values and likelihood functions for these two-stage experiments is exactly the same as it is for conventional fixed sample size stopping rules, and it can be concluded that the evidential meaning of the P-value, like that of the likelihood function, is independent of the stopping rules, as long as the P-value is not `corrected' or `adjusted'.

That result is interesting from a theoretical point of view, but it also has the practical consequence that experiments can be conducted sequentially without the need for complicated and punitive `corrections' of the P-values, as long as the P-values are correctly interpreted as a summary of evidence rather than erroneously assumed to be error rates.

\begin{quote}
The scientist who carefully examines and interprets his observations (via likelihood functions) as they arise is behaving appropriately. He is not weakening or damaging the statistical evidence, he is not spending or using up statistical capital, he should not be chastized for peeking at the data, and there is no valid basis for levying a statistical fine on his study.		\hfill  [50 p. 138]
\end{quote}

Arguments that P-values do not offer a consistent scale of evidence are often based on the false premise that evidence is one-dimensional or mistakenly assume that P-values should be `corrected' for the stopping rules.

\subsection{P-values are the product of a logically flawed approach}

\subsubsection{Fisher's disjunction is false}

An important claim that P-values are logically flawed as indices of evidence comes from a criticism of Fisher's disjunction which says that P-values do not cast doubt on the null hypothesis in the manner that Fisher suggested. Cohen illustrates that claim by drawing an analogy between Fisher's disjunction and this syllogism [51]:

\begin{quote}
If a person is an American, then he is probably not a member of Congress.\\
This person is a member of Congress.\\
Therefore, he is probably not an American.

\end{quote}
\noindent As Cohen says, the last line of that syllogism about the American is false even though it would be true if the word `probably' were omitted from the first and last lines. However, Cohen is incorrect in suggesting that it is functionally analogous to Fisher's disjunction. As Hagen pointed out in a response published a few years after Cohen's paper [52], the null hypothesis in Fisher's disjunction refers to the population, whereas in Cohen's syllogism it refers to the sample. 

Fisher's disjunction looks like this when put into the form of a syllogism:

\begin{quote}

Extreme P-values from random samples are rare under the null hypothesis.\\
An extreme P-value has been observed.\\
(Therefore, either a rare event has occurred or the null hypothesis is false.)\\
Therefore, the null hypothesis is probably false.
\end{quote}

\noindent There is nothing wrong with that, although the line in parentheses is not logically necessary. When Cohen's syllogism is altered to refer to the population, it also is true:

\begin{quote}
Members of Congress are rare in the population of Americans.\\
This person is a member of Congress.\\
(Therefore, either a rare event has occurred or this person is not a random sample from the population of Americans.)\\
Therefore, this person is probably not a random sample from the population of Americans.
\end{quote}

\noindent If a selected person turns out to be a member of Congress then an unusual event has occurred, or the person is a member of a non-American population in which members of Congress are more common, or the selection was not random. Assuming that all members of the American Congress are American there is no relevant non-American population from which the person might have been randomly selected, so the observation casts doubt on the random selection aspect. Cohen is incorrect in his assertion that the Fisher's disjunction lacks logical integrity.

(It is worth noting, parenthetically, that Cohen's paper contains many criticisms of null hypothesis testing that refer to problems arising from the use of what he describes as ``mechanical dichotomous decisions around the sacred .05 criterion''. He is correct in that, but the criticisms do not directly apply to P-values used as indices of evidence.)

\subsubsection{P-values depend on data that were not observed}

The notion that a P-value depends on data that have not been observed is an interesting idea that comes from the fact that P-values are tail areas of the sampling distribution. They sum the probability of observations \textit{at least as extreme} as that observed---those exactly as extreme as the observation and those more extreme. The more extreme observations are the observations that have not been observed. A widely quoted passage by Jeffreys says: 

\begin{quote}
What the use of P implies, therefore, is that a hypothesis that may be true may be rejected because it has not predicted observable results that have not occurred. That seems a remarkable procedure.	\hfill [53 p. 385]
\end{quote}

Such descriptions of this issue might lead to alarm, but a P-value depends on unobserved data only because the P-value expresses the extremeness of the observed data. Any useful index of extremeness has to refer in some manner to the range of the population. For instance, a friend of the author, G, is taller than most men. Let's say that he is at the 99th percentile of heights for adult men in Australia. Presumably no-one is concerned that the statements about G reference unobserved data. However, when I re-state the information in a probabilistic manner it becomes clear that it is dependent on unobserved data in exactly the same way as P-values are: the probability that a randomly selected man from Australia is taller than G is one percent; the probability that a random sample yields a P-value smaller than 0.01 when the null hypothesis is true is one percent. That type of dependency on unobserved data doesn't make the claims false or in some special way misleading, just that they are on relative scales. 

P-values are quantitative claims of extremeness of data on a relative scale. The fact that P-values are derived from tail areas of the sampling distribution in no way disqualifies them as indices of evidence.

\subsubsection{The null hypothesis is often known to be false before the experiment}

An interesting point often raised in arguments against the use of hypothesis tests and significance tests for scientific investigations is that null hypotheses are usually known to be false before an experiment is conducted. When defined as the probability under the null hypothesis of obtaining data at least as extreme as those observed, P-values would seem to be susceptible to the criticism in that they measure the discordance between the data and something that is known to be false. The argument may have some relevance to hypothesis tests, but it is irrelevant to any use of P-values in estimation and in the assessment of evidence because the null hypothesis serves as little more than an anchor for the calculation---a landmark in parameter space, as was discussed in section \ref{PSampleSize}. 

In contrast to hypothesis tests, significance tests and their P-values are immune to arguments that they lack utility when null hypotheses are routinely false.

\subsection{P-values overstate the strength of evidence}

\subsubsection{P is smaller than another statistical outcome}

There is an idea that P-values overstate the evidence against the null hypothesis that seems to come from several related issues. First, it is claimed that because P-values are [often] misconstrued as being an error rate they [often] overstate the true evidence against the null hypothesis [e.g. 14, 5]. The solution to that problem is obvious, and it doesn't involve discarding P-values!

A second rationale for saying that P-values overstate evidence seems to be based on the non-linear relationship between P-values and likelihoods or Bayes factors [54, 55]. The evidential meaning of a P-value is richer than just the maximal height of the likelihood function, as discussed in section \ref{PSampleSize}. However, even ignoring the other dimension(s) of evidence, in order for non-linearity to lead to some overestimation of the strength evidence it is necessary that the evidential strength \textit{as perceived by the researcher} be distorted or invalidated by that non-linearity. 
 Questions about whether researchers have a faulty perception of the relationship between P-values and their evidential meaning would seem to be accessible to empirical investigation, and the results of such a study would trump any theoretical argument.

A third basis for arguing that P-values overstate the strength of evidence is the fact that two-tailed P-values are consistently smaller than the posterior probability of the null hypothesis derived from a Bayesian analysis, often by an order of magnitude. That particular fact led to declaration of ``the irreconcilability of P values and evidence'' [2]. An easy response to this argument is that P-values do not measure the same thing as Bayesian posterior probabilities, and so any wish that they be numerically equivalent is misguided. Moreover, as others point out  [56], the large discrepancy between P-values and the Bayesian posteriors in that work was mostly a consequence of placing a large fraction of the prior probability on the null using a `spike and slab' prior. One-sided null hypotheses do not lend themselves to such spiking with prior probability because the null and non-null hypotheses have equivalent ranges of possible values and, as shown in another paper in the same issue of that journal, it turns out that one-tailed P-values are much closer in value to Bayesian posterior probabilities [57]. Thus, even if that argument were a reason to doubt the value of P-values as evidence, it would apply strongly only to two-tailed P-values. Even in that case it does not seem to be problematical because the one-to-one relationship between P and likelihood functions means that a Bayesian analysis has to agree with the P-values in every case except where the prior probability distribution has substantial weight distant from the observed outcome---after all, the posterior distribution is just a scaled product of the prior and the likelihood function.

\subsubsection{A large enough sample will always yield a small P-value}

It is true to say that a large enough sample will always yield a small P-value, but only when the null hypothesis is false, and in that case a small P-value that calls the null hypothesis into doubt is a \textit{good} thing. Hurlbert \& Lombardi call this criticism of P-values the ``fallacy of the obese $n$'' and say:

\begin{quote}
The demon of the overlarge sample: It lurks quietly in the darkness, waiting for researchers to pass by who are too focused on obtaining adequate sample sizes. If sample sizes are too large, one may be ``in danger'' of getting very low P values and establishing the sign and magnitude of even small effects with too much confidence. Oh, the horror of it all.	\hfill [19 p. 333-334]
\end{quote}

Some might feel that that response is unfair to the argument about large samples because the argument is not about correctly identifying null hypotheses as true or false. Instead, it is about the utility of identifying a null hypothesis as false when it is nearly true, when it is false by a trivially small amount. If the results of experiments were presented only as a P-value then the criticism might apply, but for many reasons results should never be presented only as a P-value, and such a deficient presentation should be quite rare. A small P-value can be potentially misleading in cases where the null hypothesis is false only in the absence of information regarding the observed effect size and sample size.

\subsection{P-values conflict with the likelihood principle}

The likelihood principle says that if two experiments yield proportional likelihood functions then they support the same inferences about the hypotheses. Conflict between the likelihood principle and the frequency or repeated sampling principle is well understood, but  given that P-values index likelihood functions it is difficult to see how they could conflict with the likelihood principle in a deep or inevitable way. In fact that alleged conflict is not intrinsic to P-values at all, but comes from `corrections' to P-values that force the P-values to conform to the frequency principle, as can be inferred from the usual illustration of the conflict, which involves a tale of two statisticians testing a coin for bias.

 Imagine that two frequentist statisticians A and B collaborate on an experiment but have not negotiated stopping rules in advance---presumably each thinks the other is a reasonable man who would have the same stopping rules in mind. However the unstated intention of A is to toss the coin six times and count the number of heads (a standard fixed sample size design) and B intends to toss the coin repeatedly until a heads comes up (a sequential design that is often called negative binomial sampling). The experimental result was five tails in a row followed by one head and so the stopping rules of each statistician are simultaneously satisfied. Statistician A calculates a P-value of 0.03 using the conventional formula for binomial sampling

\begin{equation}
\label{binomExpt}
\text{P-value} = \binom{6}{1}p(1-p)^5 + \binom{6}{0}p(1-p)^6
\end{equation}

\noindent where $p=0.5$ is the probability of the coin turning up heads under the null hypothesis. Statistician B calculated the P-value of 0.11 from the probability of needing 5 or more throws before the first heads using the formula for `negative' binomial sampling

\begin{equation}
\label{negBinomExpt}
\text{P-value} = p(1-p)^5 + p(1-p)^6 + p(1-p)^7 \dots = p(1-p)^4
\end{equation}

\noindent Statistician A claims a `significant' result because P$<\alpha=0.05$ but statistician B claims that result is `not significant'. Their behavioural inferences differ, and a fight ensues.

That example is supposed to illustrate a conflict between P-values and the likelihood principle like this: the one dataset yielded two different inferences via the P-values, but the probabilities of the observed result assumed by the two statisticians are proportional---$\binom{6}{1}p(1-p)^5$ for statistician A and $p(1-p)^5$ for statistician B---thus the likelihood functions are proportional and, according to the likelihood principle, the statisticians should both make the same inference.  Ergo, conflict. However, it is important to note that while the conflict is real, it is conflict between Neyman--Pearsonian hypothesis testing and the likelihood principle. Both statisticians in the example calculated and used their P-values as if they were long term error rates rather than as indices of evidence. The difference between equations \ref{binomExpt} and \ref{negBinomExpt} is that the latter assumes that the experiment had a sequential design and adjusts the calculation of the P-value to take the consequences of that design on error rates into account, but, as was shown in section \ref{sec:stopping_rules}, where P-values are used as indices of evidence then it is necessary not to adjust them for the sequential design of the experiment. Inference from the unadjusted P-values might differ from the behavioural inference that complies with the frequency principle, but inference from the unadjusted P-values via the likelihood functions that they index can be completely compatible with the likelihood principle.

The apparent conflict between frequentism and the likelihood principle comes not from the frequentist conception of probability or from the nature of P-values, but from the conflict between experimental evidence and the error-decision framework of Neyman and Pearson.

%

\section{Discussion}

By examining the largely unremarked relationship between P-values and likelihood functions this paper  shows how P-values can and should be used in evidence-based inference. The results come from simply documenting the emergent properties of P-values rather than from consideration of presupposed definitions, and that unusual approach allows the true nature of P-values to be discerned without interference by preconceptions regarding what P-values might be. The results show that, despite claims to the contrary, P-values do summarize the evidence in experimental data, but they summarise by indexing a likelihood function rather than by their numerical value. Thus inevitably the likelihood function provides a more complete depiction of that evidence than the P-value, not only because it shows the strength of evidence on an easily understood scale, but also because it shows how that evidence relates to a continuous range of parameter values. However, for the those understandings to become embedded into the scientific community (and, dare I say it, the statistical community) it will be necessary for likelihood functions themselves to be more widely understood. As Edwards pointed out [58], users will readily gain a feel for the practical meaning of likelihood functions, and thus P-values, simply by using them. Hacking succinctly paraphrased Edwards:

\begin{quote}
if you use likelihoods in reporting experimental work, you will get to understand them just as well as you now think you understand probabilities.	[59]

\end{quote}

Full likelihood functions give a more complete picture of the evidential meaning of experimental results than do P-values, so they are a superior tool for viewing and interpreting those results. However, it is sensible to make a distinction between the processes of drawing conclusions from experiments and displaying the results. For the latter, it is probably unnecessary and undesirable for a likelihood function be included every time a P-value might otherwise be specified in research papers. To do so would often lead to clutter and would waste space because, given knowledge of sample size and test type, a single P-value corresponds to a single likelihood function and thus stands as an unambiguous index. However, during deliberative interpretation of data and while forming scientific inference it would be sensible for an experimenter to view the likelihood functions along with the P-values. Software packages used for statistical analysis should therefore offer, by default, a likelihood function as part of their report on the result of tests of significance. Of course, for such a change to make sense it will be necessary that textbooks for introductory courses on statistics be revised both to introduce likelihood functions and to make clear the distinction between Fisherian significance tests and Neyman--Pearsonian hypothesis test. The approach used in this paper to show the interrelatedness of P-values and likelihood functions appears to have some pedagogic utility, and could serve as a useful model for textbook authors to follow. In particular, the determination of likelihood functions using both P-values and the cumulative power curves should allow presentation of several important concepts in an integrated manner.

The P-value from a significance test has properties consistent with it being a useful tool for scientific inference. Its widespread use for such purposes is therefore neither an unfortunate accident of history, nor is it a consequence of institutionalization of a faulty approach, despite such claims in the literature. The conventional usage of P-values is not free from the need for reform, however, but efforts by statisticians to improve the manner in which scientific experiments are analyzed and interpreted should be directed at increasing explicit consideration of the evidence provided by the experimental data. Presentation of the statistical analyses of experimental results should always be accompanied by what Abelson called `principled argument' [47] and ideally by replication of the key experiments. Both of those are better supported by evidential assessment of the experimental results than by error rates.

\section{References}
\begin{enumerate}
\item
Hubbard R, Lindsay RM (2008) Why P Values Are Not a Useful Measure of Evidence in Statistical Significance Testing. Theory \& Psychology 18:69--88.

\item	Berger J, Sellke T (1987) Testing a Point Null Hypothesis: The Irreconcilability of P Values and Evidence. Journal of the American Statistical Association 82:112--122.

\item	Cumming G (2008) Replication and p intervals: p values predict the future only vaguely, but confidence intervals do much better. Perspectives on Psychological Science 3:286--300.

\item	Johansson T (2010) Hail the impossible: p-values, evidence, and likelihood. Scandinavian Journal of Psychology 52:113--125.

\item	Hubbard R, Bayarri MJ (2003) Confusion over measures of evidence (p’s) versus errors ($\alpha$’s) in classical statistical testing. The American Statistician 57:171--178.

\item	Schervish MJ (1996) P values: what they are and what they are not. The American Statistician 50:203--206.

\item	Gill J (1999) The Insignificance of Null Hypothesis Significance Testing. Political Research Quarterly 52:647--674.

\item	Lambdin C (2012) Significance tests as sorcery: Science is empirical--significance tests are not. Theory \& Psychology 22:67--90.

\item	Lecoutre B, Lecoutre M-P, Poitevineau J (2001) Uses, Abuses and Misuses of Significance Tests in the Scientific Community: Won’t the Bayesian Choice Be Unavoidable? International Statistical Review / Revue Internationale de Statistique 69:399--417.

\item	Lecoutre M-P, Poitevineau J, Lecoutre B (2003) Even statisticians are not immune to misinterpretations of Null Hypothesis Significance Tests. International Journal of Psychology 38:37--45.

\item	Lew MJ (2012) Bad statistical practice in pharmacology (and other basic biomedical disciplines): you probably don’t know P. British Journal of Pharmacology 166:1559--1567.

\item	Wagenmakers E-J (2007) A practical solution to the pervasive problems of p values. Psychonomic Bulletin \& Review 14:779--804.

\item	Jaynes ET (1980) What is the question?, in \textit{Bayesian Statistics}, eds Bernardo JM, deGroot MH, Lindley DV, Smith AFM (Valencia University Press, Valencia), 

\item	Berger J (2003) Could Fisher, Jeffreys and Neyman Have Agreed on Testing? Statistical Science 18:1--12.

\item	Schweder T, Norberg R (1988) A Significance Version of the Basic Neyman--Pearson Theory for Scientific Hypothesis Testing. Scandinavian Journal of Statistics 15:225--242.

\item	Mayo DG, Spanos A (2011) Error statistics, in Philosophy of statistics, eds Bandyopadhyay PS \& Forster MR. Elsevier, pp 1--46.

\item	Gigerenzer G (1992) The superego, the ego, and the id in statistical reasoning, in \textit{A Handbook for Data Analysis in the Behavioral Sciences: Methodological Issues}, eds Gideon K, Lewis C (L. Erlbaum Associates, Hillsdale, NJ), pp 311--339.

\item	Gigerenzer G (1998) We need statistical thinking, not statistical rituals. Behavioral and Brain Sciences 21:199--200.

\item	Hurlbert SH, Lombardi CM (2009) Final collapse of the Neyman--Pearson decision theoretic framework and rise of the neoFisherian. Annales Zoologici Fennici 46:311--349.

\item	Oakes M (1986) Statistical inference (Wiley, Chichester; New York).

\item	Haller H, Krauss S (2002) Misinterpretations of significance: A problem students share with their teachers. Methods of Psychological Research 7:1--20.

\item	Halpin PF, Stam HJ (2006) Inductive inference or inductive behavior: Fisher and Neyman--Pearson approaches to statistical testing in psychological research (1940-1960). The American Journal of Psychology 119:625--653.

\item	Huberty CJ (1993) Historical origins of statistical testing practices: The treatment of Fisher versus Neyman--Pearson views in textbooks. The Journal of Experimental Educational 61:317--333.

\item	Fisher RA (1966) The design of experiments (London Oliver \& Boyd, Edinburgh).

\item	Neyman J, Pearson E (1933) On the Problem of the Most Efficient Tests of Statistical Hypotheses. Philosophical Transactions of the Royal Society of London Series A 231:289--337.

\item	Neyman J (1957) ``Inductive Behavior'' as a Basic Concept of Philosophy of Science. Revue de l'Institut International de Statistique 25:7--22.

\item	Strasak A, Zaman Q, Marinell G, Pfeiffer K (2007) The Use of Statistics in Medical Research: A Comparison of The New England Journal of Medicine and Nature Medicine. The American Statistician 61:47--55.

\item	Hacking I (1965) Logic of statistical inference. (Cambridge University Press, Cambridge).

\item	Royall R (1997) Statistical evidence: a likelihood paradigm (Chapman \& Hall/CRC)

\item	Edwards AWF (1992) Likelihood (The Johns Hopkins University Press, Baltimore, MD).

\item	Sackrowitz H, Samuel-Cahn E (1999) P Values as Random Variables---Expected P Values. American Statistician 53: 326--331.

\item	Murdoch DJ, Tsai Y-L, Adcock J (2008) P-Values are Random Variables. The American Statistician 62:242--245.

\item	Bhattacharya B, Habtzghi D (2002) Median of the P-Value Under the Alternative Hypothesis. The American Statistician 56:202--206.

\item	Hoenig J, Heisey D (2001) The Abuse of Power: The Pervasive Fallacy of Power Calculations for Data Analysis. The American Statistician 55: 1--6

\item	Royall R (2000) On the Probability of Observing Misleading Statistical Evidence. Journal of the American Statistical Association 95:760--768.

\item	Lombardi CM, Hurlbert SH (2009) Misprescription and misuse of one-tailed tests. Austral Ecology 34:447--468.

\item	Bland JM, Bland DG (1994) Statistics Notes: One and two sided tests of significance. BMJ 309:248--248.

\item	Kaiser HF (1960) Directional Statistical Decisions. Psychological Review 67:160--167.

\item	Rice WR, Gaines SD (1994) `Heads I win, tails you lose': testing directional alternative hypotheses in ecological and evolutionary research. Trends in Ecology \& Evolution 9:235--237.

\item	Dubey SD (1991) Some thoughts on the one-sided and two-sided tests. Journal of Biopharmaceutical Statistics 1:139--150.

\item	Ringwalt C, Paschall MJ, Gorman D, Derzon J, Kinlaw A (2011) The Use of One- Versus Two-Tailed Tests to Evaluate Prevention Programs. Evaluation and the Health Professions 34:135--150.

\item	Eysenck HJ (1960) The concept of statistical significance and the controversy about one-tailed tests. Psychological Review 67:269--271.

\item	Student (1908) The Probable Error of a Mean. Biometrika 6:1--25.

\item	Freedman LS (2008) An analysis of the controversy over classical one-sided tests. Clinical Trials 5:635--640.

\item	Burke CJ (1953) A brief note on one-tailed tests. Psychological Bulletin 50:384.

\item	Gigerenzer G (2004) Mindless statistics. Journal of Socio-Economics 33:587--606.

\item	Abelson RP (1995) Statistics as principled argument (Taylor \& Francis, Hillsdale, NJ).

\item	Royall RM (1986) The Effect of Sample Size on the Meaning of Significance Tests. The American Statistician 40:313--315.

\item	Fisher R (1955) Statistical Methods and Scientific Induction. Journal of the Royal Statistical Society Series B (Methodological) 17:69--78.

\item	Royall R (2004) The likelihood paradigm for statistical evidence, in \textit{The nature of scientific evidence: statistical, philosophical and empirical considerations}, eds Taper ML \& Lele SR (University Of Chicago Press, pp 119--152.

\item	Cohen J (1994) The earth is round (p <.05). American Psychologist 49:997--1003.

\item	Hagen RL (1997) In praise of the null hypothesis statistical test. American Psychologist 52:15--24.

\item	Jeffreys H (1961) Theory of probability (Oxford University Press, Oxford).

\item	Goodman SN (1993) P values, hypothesis tests, and likelihood: implications for epidemiology of a neglected historical debate. American Journal of Epidemiology 137:485--496.

\item	Goodman SN (2001) Of P-Values and Bayes: A Modest Proposal. Epidemiology 12:295--297.

\item	Vardeman SB (1987) Testing a Point Null Hypothesis: The Irreconcilability of P Values and Evidence: Comment. Journal of the American Statistical Association 82:130--131.

\item	Casella G, Berger RL (1987) Reconciling Bayesian and frequentist evidence in the one-sided testing problem. Journal of the American Statistical Association 82:106--111.

\item	Edwards AFW (1972) Likelihood: an account of the statistical concept of likelihood and its application to scientific inference (Cambridge University Press, Cambridge).

\item	Hacking I (1972) Likelihood. British Journal for the Philosophy of Science 23:132--137.

\end{enumerate}

\section{Appendix}
R code for the likelihood functions based on P-values:
\begin{verbatim}
LikeFromStudentsTP<-function(n,x,Pobs,test.type){
	# test.type can be 'one.sample', 'two.sample' or 'paired'
	# n is the sample size (per group for test.type = 'two.sample')
	# Pobs is the observed P-value
	# h is a small number used in the trivial differentiation
	h<-10^-7
	PowerDn<-power.t.test('n'=n, 'delta'=x, 'sd'=1,
		'sig.level' = Pobs-h,  'type'= test.type, 'alternative'='one.sided')
	PowerUp<-power.t.test('n'=n, 'delta'=deltaOnSigma, 'sd'=1,
		'sig.level' = Pobs+h,  'type'= test.type, 'alternative'='one.sided')
	PowerSlope<-(PowerUp$power-PowerDn$power)/(h*2)
	L<-PowerSlope
}

# run an example
n<-10
P<-0.025
deltaOnSigma <- 0.01*c(-100:500) #-1 to 5 to catch most of the detail needed
type<-'two.sample'
y<-LikeFromStudentsTP(n,deltaOnSigma,P,type)
plot(deltaOnSigma,y)

\end{verbatim}

\noindent R code for P-value probability density functions:
\begin{verbatim}
# Function pcvs returns a vertical section through P-value cloud.
# Equivalent to the probability distribution of the p-value.

require(numDeriv)

#cumulative distribution
pcvsCum <- function(x)	{
			y<-power.t.test('n'=n, 'delta'=DeltaOnSigma, 'sd'=1, 'sig.level'=x, 'type'=test.type, 				'alternative'=tails)
			y$power
						}
						
#probability density distribution
pcvs <-function(x) {
	grad(pcvsCum,x) #function in numDeriv package
	}
	
# run example	
DeltaOnSigma <- 0.5
n<-10
test.type<-'two.sample'
tails<-'one.sided'
x1<-0.01*c(1:99)
x <- c(0.001,x1,0.999)
y<-pcvs(x)
plot(x,y)

\end{verbatim}

\end{document}